# $Ba_{1-x}Na_xTi_2Sb_2O$ (0.0 ≤ x≤ 0.33): A Layered Titanium-based Pnictide Oxide Superconductor.

Phuong Doan[†,Π], Melissa Gooch[‡,Π], Zhongjia Tang[†,Π], Bernd Lorenz[‡,Π], Angela Möller [†,Π], Joshua Tapp[†,Π], Paul C.W. Chu[‡,Π,§] and Arnold M. Guloy[*,†,Π].

[†]Department of Chemistry, [‡]Department of Physics, and [Π]Texas Center for Superconductivity, University of Houston, Houston, TX 77204; [§]Lawrence Berkeley National Laboratory, 1 Cyclotron Road, Berkeley, CA 94720.

**ABSTRACT:** A new layered Ti-based pnictide oxide superconductor, $Ba_{1-x}Na_xTi_2Sb_2O$ (0.0 ≤ x≤ 0.33), is reported. X-ray studies reveal it crystallizes in the tetragonal $CeCr_2Si_2C$ structure. The undoped parent compound, $BaTi_2Sb_2O$ (P4/mmm; $a$=4.1196(1)Å; $c$=8.0951(2)Å), exhibits a CDW/SDW transition at 54K. Upon chemical doping with Na, the CDW/SDW transition is systematically suppressed and superconductivity arises with the critical temperatures, Tc, increasing to 5.5 K. Bulk superconductivity is confirmed by resistivity, magnetic and heat capacity measurements. Like the high-Tc cuprates and the iron pnictides, superconductivity in $BaTi_2Sb_2O$ arises from an ordered state. Similarities and differences to the cuprate and iron pnictide superconductors are discussed.

The discovery of superconductivity in iron pnictides and chalcogenides had sparked an immense activity in the field of high-temperature superconductivity and revitalized the interest in layered transition metal oxides, chalcogenides, and pnictides.[1] The proximity of spin and/or charge ordering in intrinsic compounds to a superconducting ground state emerging through carrier doping appears to be a common feature of unconventional superconductors, e.g. cuprates and heavy fermions (antiferromagnetism), iron pnictides (spin density wave, SDW), and layered chalcogenides (charge density wave, CDW).[2,3,4] Low-dimensional metallic systems are susceptible to electronic instabilities, such as CDW or SDW formation, which occur when their Fermi surfaces are nested.[5] Nesting occurs when large areas of the Fermi surface are nearly parallel, i.e. only separated by a characteristic vector **q** in reciprocal space, that leads to an instability with a long range order of spins or charges. The competition and mutual interaction of these different orderings in strongly correlated systems is a foundation for novel physics (e.g. spin fluctuation mediated superconductivity[6], quantum critical phenomena[7]), and provides new opportunities for exploratory materials synthesis.

Based on the common intrinsic behavior with the cuprate and iron pnictide superconductors, the titanium pnictide oxides, $Na_2Ti_2Pn_2O$ (Pn = As, Sb)[8], have been proposed as possible candidates for new superconductors.[9,10] These metallic compounds are members of a wider family of layered transition metal pnictide oxides,[10,11] and they exhibit distinct phase transitions[12] that were associated with the formation of a charge- or spin-density waves.[13-15] The possible origin of this transition have been discussed,[16] where band structure calculations reveal a nesting feature in the Fermi surface that was later found to be also associated with a periodic lattice distortion.[17]

Similar to the layered structures and magnetic behavior of the iron pnictides, where alkali and alkaline earth metals form stacked layered structures with $Fe_2As_2$ slabs,[18,19] $BaTi_2As_2O$, an alkaline earth metal analog to $Na_2Ti_2Pn_2O$ was recently reported.[20] It features magnetic and resistivity anomalies similar to the transitions observed in $Na_2Ti_2Pn_2O$. However, attempts to induce superconductivity in both $Na_2Ti_2Pn_2O$ and $BaTi_2As_2O$ by chemical substitution and Li-intercalation have so far been unsuccessful, albeit lower CDW/SDW transition temperatures were observed.[20]

Herein we report the synthesis of a layered titanium pnictide oxide, $BaTi_2Sb_2O$, that exhibits a CDW/SDW transition ($T_s$ = 54K), albeit at a lower temperature. Partial substitution of Ba by Na, results in a superconducting phase, $Ba_{1-x}Na_xTi_2Sb_2O$ (0.05 ≤ x ≤ 0.33), with maximum $T_c$ of 5.5 K.

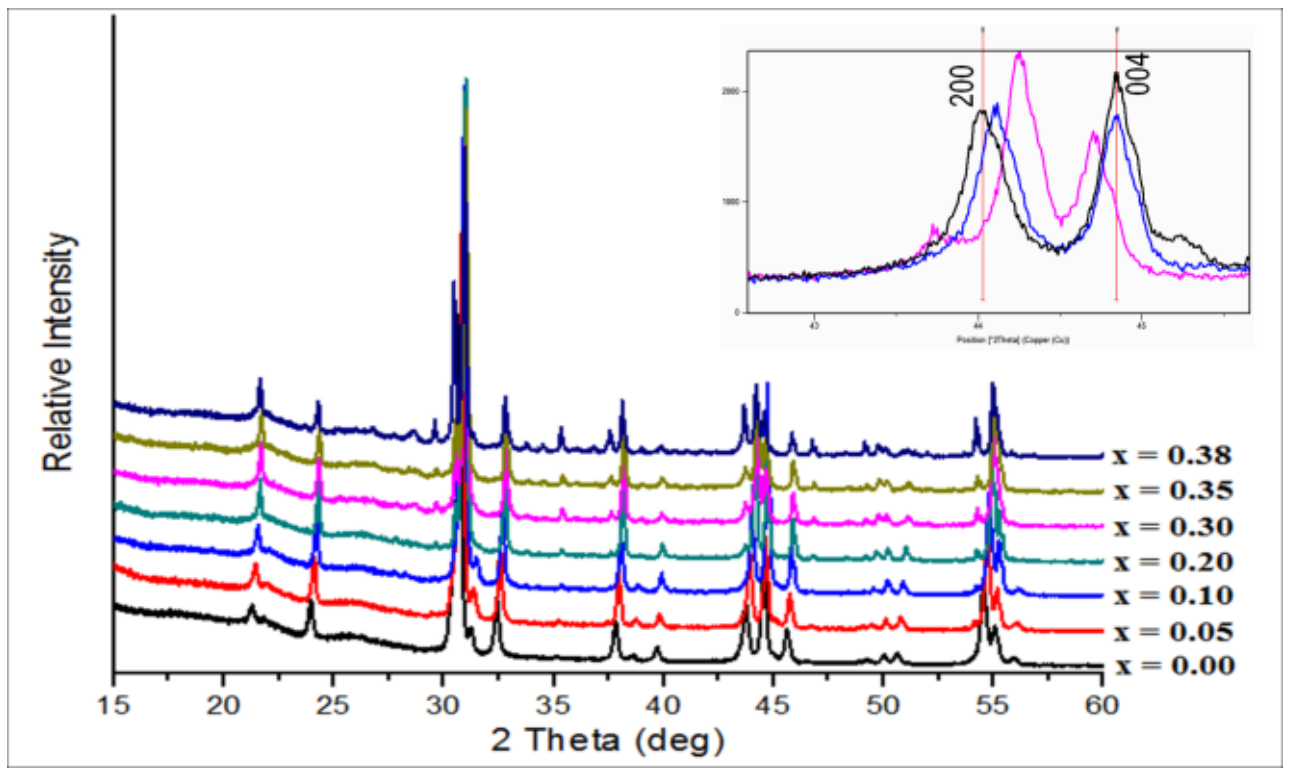

***Figure 1.*** XRPD of $Ba_{(1-x)}Na_xTi_2Sb_2O$ with x= 0.00-0.38. The background at low angles arises from the scattering of the Mylar protective film. The inset shows the [200] and [004] diffraction peaks for representative samples with x= 0.00 (black), 0.10 (blue), and 0.30(pink).

The title compounds were synthesized by the high-temperature reactions of the appropriate reactants (BaO (99.5%, STREM), $BaO_2$ (95%, Sigma Aldrich); $Na_2O$ (80% $Na_2O$ and 20% $Na_2O_2$, Sigma Aldrich); Ti (99.99%, Sigma Aldrich), Sb (pieces, 99.999%, Alfa Aesar) in welded Nb containers within evacuated quartz jackets.

The reactions were performed at 900ºC for 3 days, and slow cooled (2ºC/min) to 200 ºC. Additional regrinding and sintering at 900 ºC for another 3 days was performed to ensure phase homogeneity. The air and moisture sensitive polycrystalline products had a dark gray color. All experimental handling and physical measurements (transport, magnetic and heat capacity) were performed under inert conditions within a purified Ar-atmosphere glove box, with total $H_2O$ and $O_2$ levels < 0.1 ppm. Details of the experimental procedures and measurements, results of the chemical analyses, X-ray crystallographic data and tables, and list of lattice parameters are given in the Supporting Information.

The phase purity of the resulting polycrystalline samples were investigated by X-ray powder diffraction. The XRPD data (Figure 1) of $BaTi_2Sb_2O$ ($a$=4.1196(1)Å; $c$=8.0951(2)Å) and $Ba_{1-x}Na_xTi_2Sb_2O$ were refined to the $BaTi_2As_2O$ and $CeCr_2Si_2C$ models (P4/mmm) by the Rietveld method, using the Rietica program.[21] Phase analyses of the XRPD of $Ba_{(1-x)}Na_xTi_2Sb_2O$ with x = 0.00-0.38 indicates a homogeneity range exists up to about~30-35% Na doping levels, with the formation of $Na_2Ti_2Sb_2O$ occurring for x>0.30. To establish the actual Na doping levels, elemental analyses on the bulk and on several crystallites of each phase pure Na-doped polycrystalline samples were carried out using an inductively coupled plasma/mass spectrometer (ICPMS), using laser ablation. Results show the molar ratio Ba : Na were consistent with the nominal composition, with the Na limit of x = 0.33(2). Moreover, the composition of the impurities for x>0.30 were also established to be $Na_2Ti_2Sb_2O$. No contamination from the container (Nb) was observed in any of the samples. Further analyses of the lattice parameters also show a systematic change with increasing Na content. This is clearly reflected by the shift of the [200] peaks to higher 2θ, with increasing Na content, while the [004] peaks shift to the lower values (Fig. 1, inset). This is consistent with a contraction along *a,* and concomitant with an elongation along *c*.

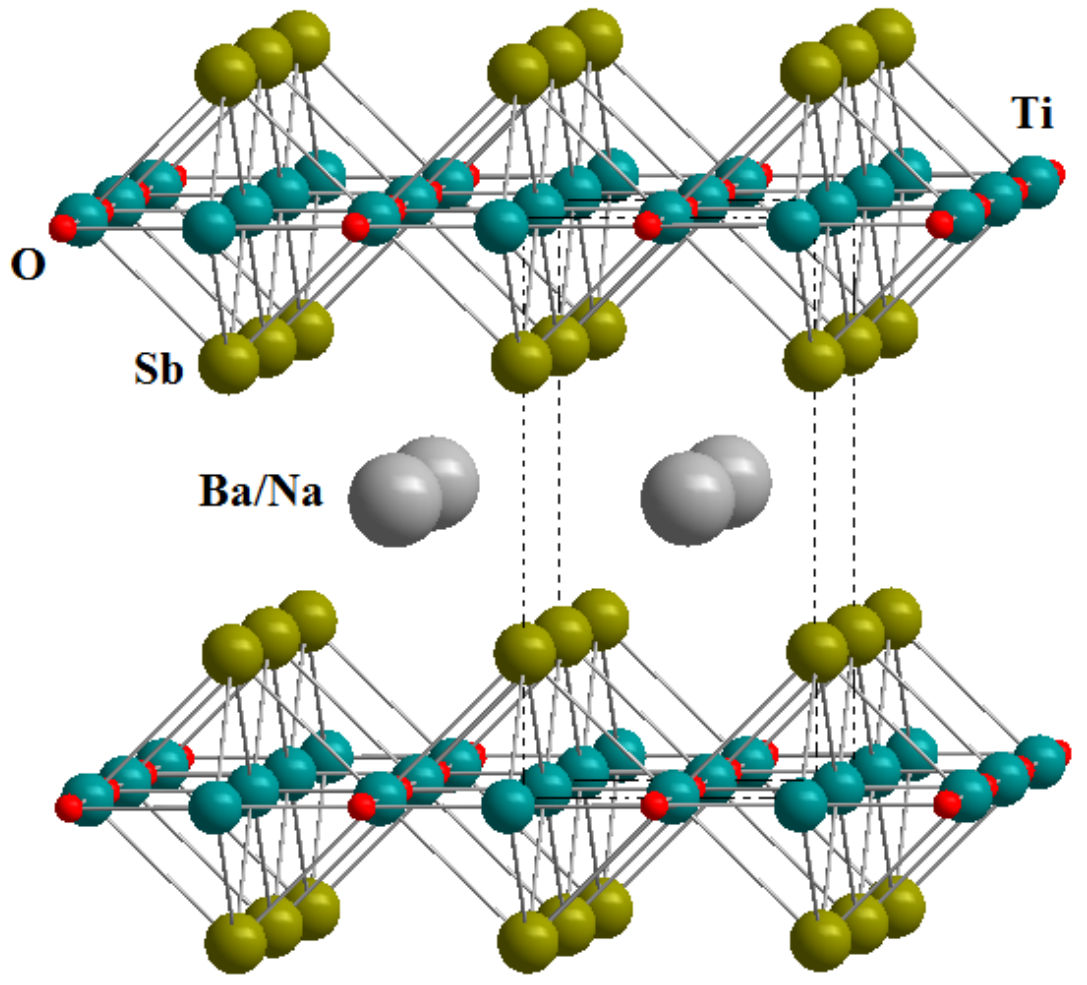

***Figure 2.*** Crystal structure of $BaTi_2Sb_2O$. Ba, Ti, Sb and O are shown as grey, blue, green, and red spheres, respectively.

Like $BaTi_2As_2O$ the title compound crystallizes in the $CeCr_2Si_2C$ structure type, a filled variant of the $CeMg_2Si_2$-type.[22] The crystal structure as shown in Figure 2, features Ti and O atoms that form $OTi_{4/2}$ anti-perovskite layers, analogous to the $CuO_2$ layers in the HTS cuprates. These layers are then capped by Sb atoms above and below the empty $Ti_4$ squares forming $Ti_2Sb_2O$ slabs that are stacked alternately with layers of Ba/Na atoms along the *c*-axis. The $Ti_2Sb_2O$ slabs are similar to $Ti_2Pn_2O$ layers in $Na_2Ti_2Pn_2O$. The Ba/Na atoms lie above and below the O atoms to form an octahedral $Ti_4(Ba/Na)_2$ coordination around oxygen. Each Ti is also octahedrally coordinated by 2 O ($d_{Ti\text{-}O}$ = 2.054(1) Å) and 4 Sb atoms ($d_{Ti\text{-}Sb}$ = 2.872(1) Å. The network of $OTi_{4/2}(Ba/Na)_{2/2}$ octahedra is analogous, but inverse, to a tetragonal perovskite ($ReO_3$-type), with 2 Sb atoms occupying cuboctahedral cavities of the anti-perovskite network. However, unlike the isotypical silicides ($CeCr_2Si_2C$-type) having $Si_2$ units with short Si interlayer distances, Sb-Sb distances in $BaTi_2Sb_2O$ are essentially non-bonding ($d_{Sb\text{-}Sb}$ > 4.0 Å). Upon Ba substitution with Na, a further expansion of the interlayer distance (c/2) is observed, accompanied by a slight contraction of the Ti-Ti distances in the $Ti_2O$ sheets.

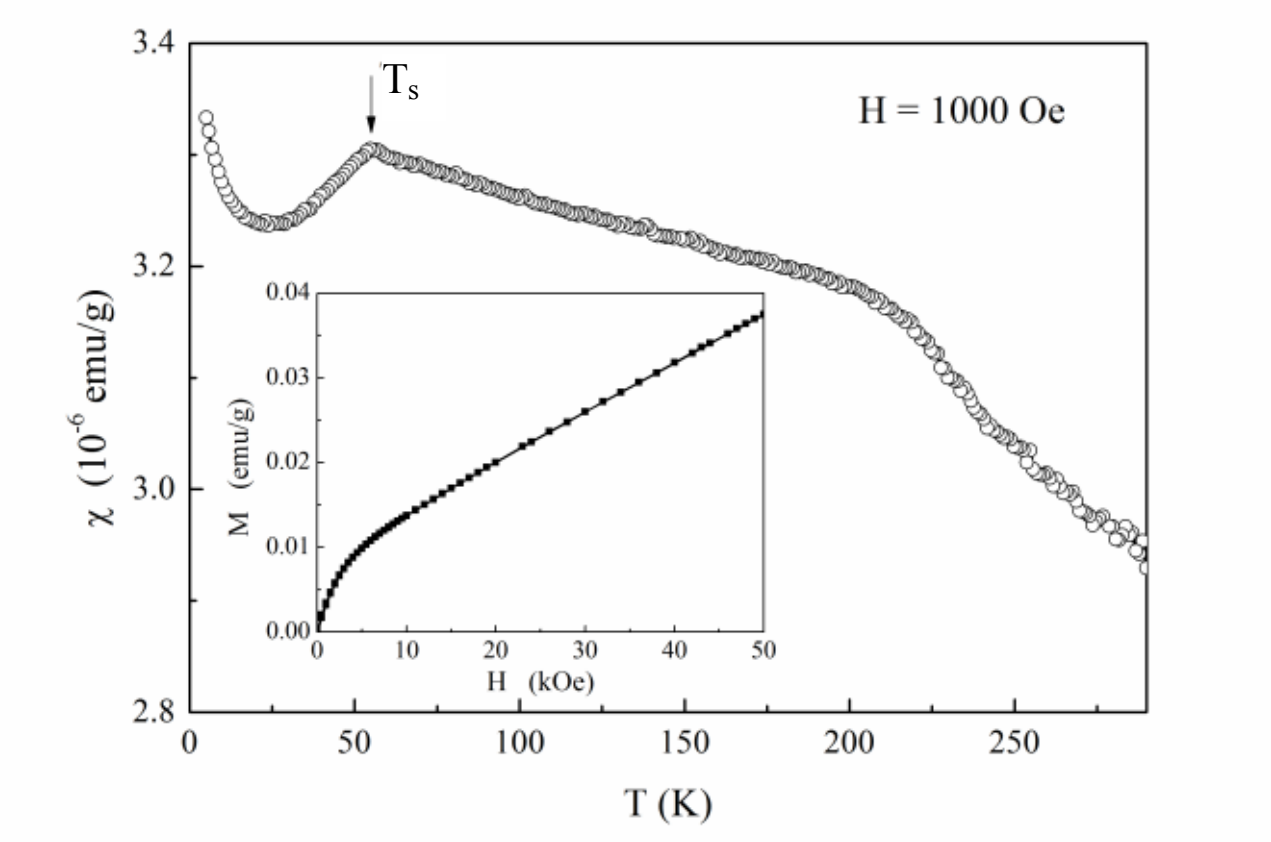

***Figure 3.*** Magnetic susceptibility of $BaTi_2Sb_2O$. Inset: Field dependence of magnetization.

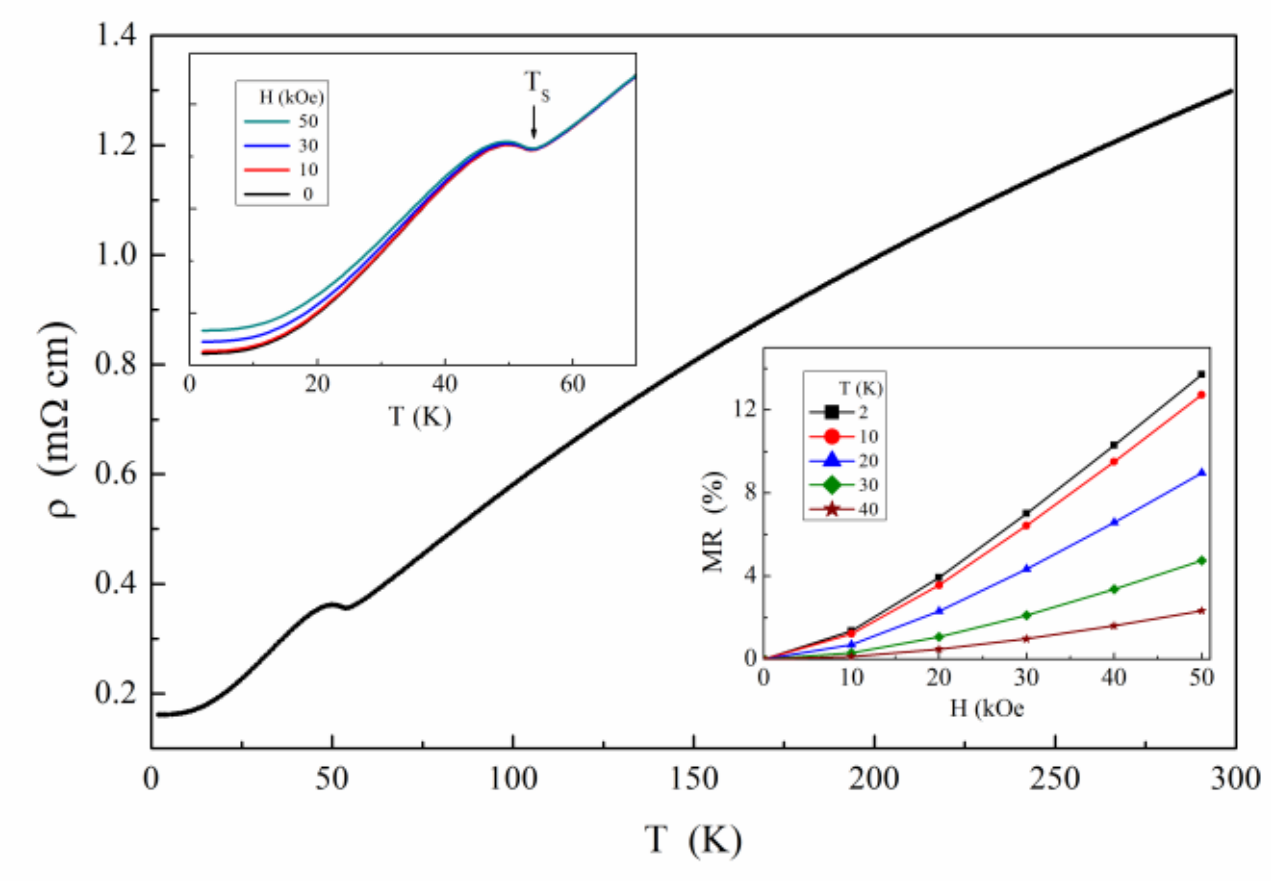

***Figure 4.*** Resistivity and magnetoresistance (MR) of $BaTi_2Sb_2O$.

The magnetic susceptibility of $BaTi_2Sb_2O$ is shown in Figure 3 clearly reveals a sharp drop at $T_s$ = 54, K indicating the onset of a magnetic order. Although the temperature dependence of the susceptibility is qualitatively similar to the data recently obtained for the As-based compound, $BaTi_2As_2O$, the magnetic transition temperature of $BaTi_2Sb_2O$ is remarkably lower (~ 25 % of the value reported for $BaTi_2As_2O$).[20] A similar trend of lower magnetic $T_s$ with increasing pnictide ionic radii (As → Sb) was also observed in the Na-based compounds, $Na_2Ti_2Pn_2O$, with $T_s$ = 320 K (Pn = As) and 120 K (Pn = Sb).[14] The nearly linear T-dependence of the susceptibility above $T_s$ is consistent with the data for $BaTi_2As_2O$ and $Na_2Ti_2Pn_2O$. There is an apparent slope change of χ(T) near 220 K, the origin of which has yet to be investigated.

The resistivity of $BaTi_2Sb_2O$ at zero fields and in magnetic fields up to 50 kOe is shown in Figure 4 (main panel and upper inset). The sharp slope change at $T_S$ = 54 K followed by a hump is a clear indication of a CDW/SDW transition, similar to that reported for $Na_2Ti_2Sb_2O$ and $BaTi_2As_2O$.[17,20] Resistivity in magnetic fields of 50 kOe increases by 14 % at low temperatures (Fig. 4, lower inset). What distinguishes $BaTi_2Sb_2O$ from the other pnictide oxides are the significantly lower $T_S$, and a much larger magnetoresistance (MR) that exceeds the reported value for $BaTi_2As_2O$ at 50 kOe by a factor of 5.[20] This indicates strong coupling between the SDW state and the charge carriers leading to a lower energy scale of the CDW/SDW state and a higher sensitivity of the low-temperature state to magnetic fields. Thus, the possibility that superconductivity could be induced with proper doping is enhanced, similar to the pnictide and chalcogenide superconductors.

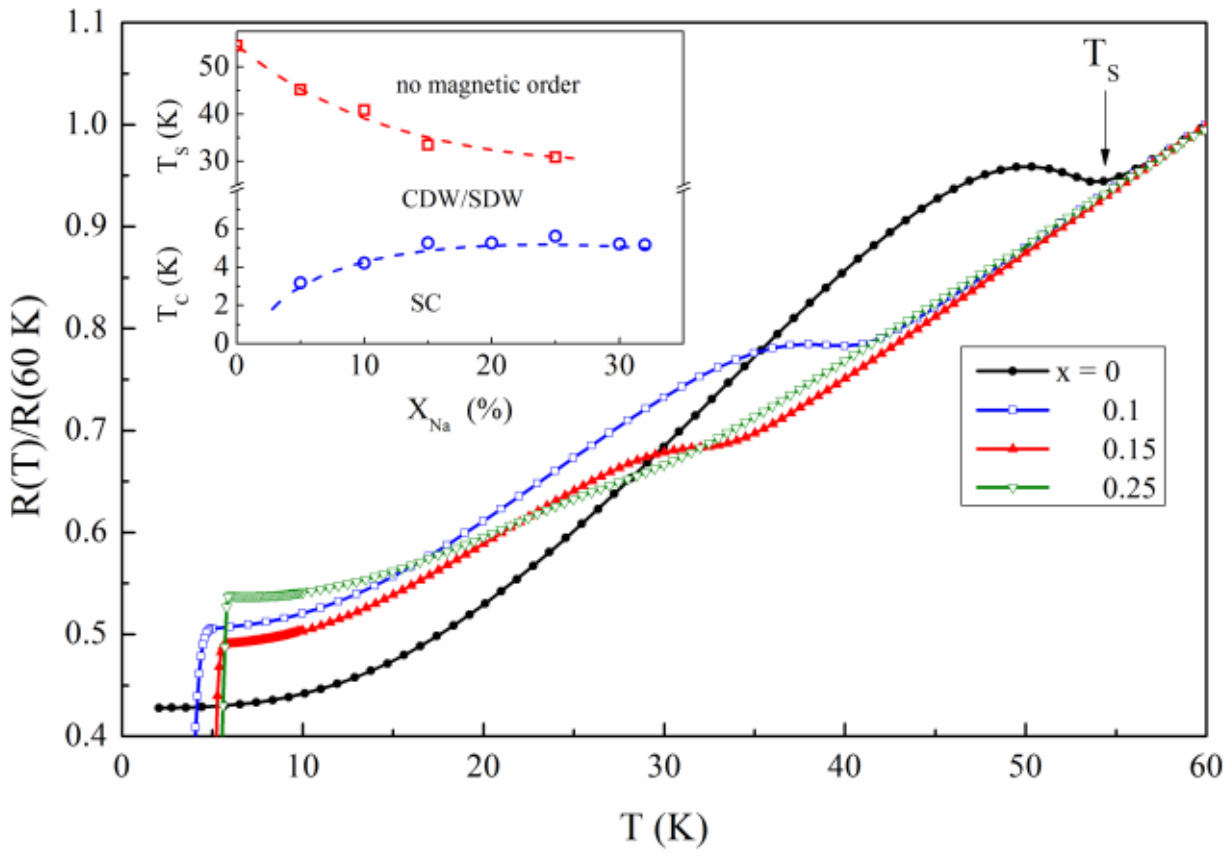

***Figure 5.*** Normalized resistance of $Ba_{1-x}Na_xTi_2Sb_2O$ near the CDW/SDW transition. The inset shows the phase diagram derived from resistivity and magnetization measurements.

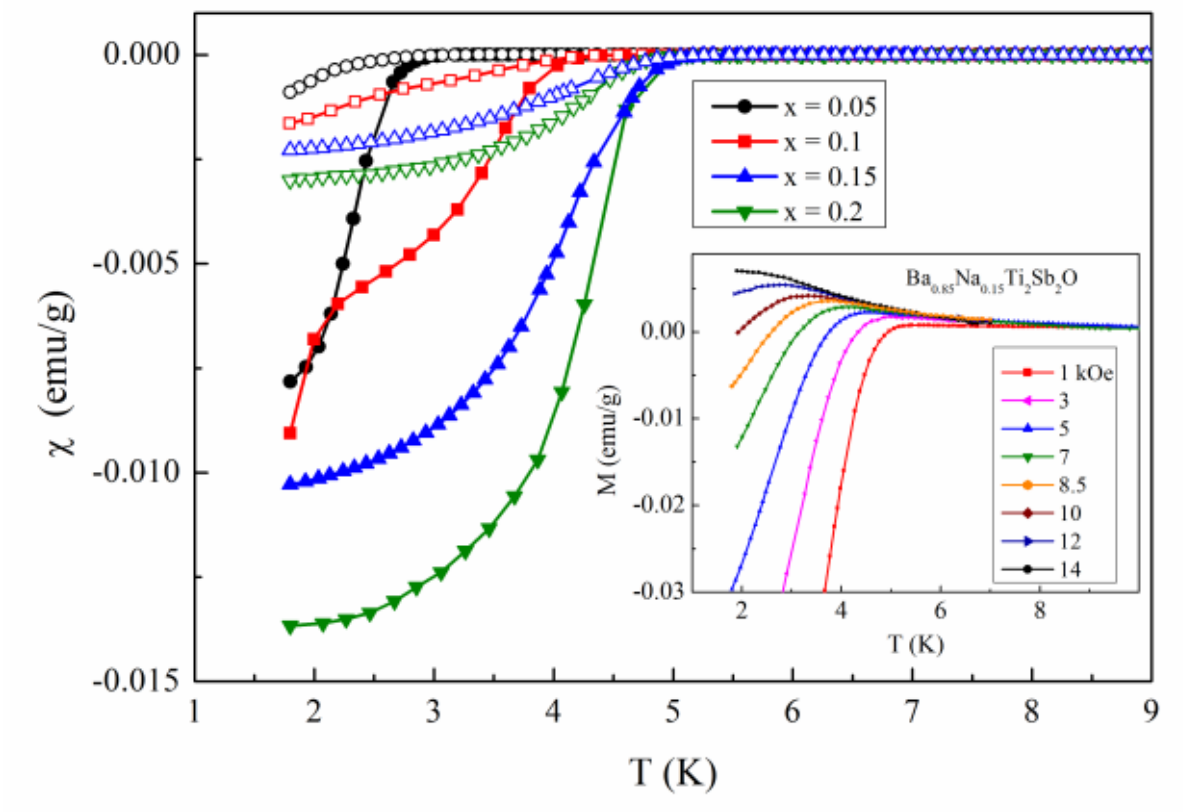

***Figure 6.*** Low-temperature magnetic susceptibility of $Ba_{1-x}Na_xTi_2Sb_2O$ for selected values of x. Open and closed symbols refer to FC and ZFC data, respectively. Inset shows the field dependence of the magnetization at 5 K for x = 0.15 under FC conditions.

Substitution of Ba by Na in $BaTi_2Sb_2O$ corresponds to a 'hole' (p-type) doping of the $Ti_2Sb_2O$-layers. As shown in Figure 5, the CDW/SDW transition ($T_S$) decreases, with increasing Na content (x), from 54 K (x=0) to 30 K (x=0.25). Also, $T_S$ does not extrapolate to zero up to a Na solubility limit of about 33% although the size of the hump-like anomaly does significantly decrease with x. More importantly, a sharp drop of the resistivity to zero indicates the emergence of superconductivity at lower temperatures. The superconducting phase diagram together with the CDW/SDW transition line is summarized in Figure 5 (inset).

The superconducting state of $Ba_{1-x}Na_xTi_2Sb_2O$ is confirmed by a diamagnetic signal, the onset of which is used to define the superconducting $T_c$. Figure 6 shows the zero-field cooled (ZFC) and field-cooled (FC) magnetization data for selected Na-doped samples, measured at 10 Oe. $T_c$ increases rapidly between x=0.05 and x=0.15, and plateaus at higher x values. The diamagnetic signal also increases and reaches its maximum at x = 0.15. High-field magnetization (FC) measurements for x = 0.15 (inset of Figure 6) also reveal the suppression of $T_c$ with increasing field.

The bulk nature of the superconducting transition in $Ba_{1-x}Na_xTi_2Sb_2O$ is confirmed by a distinct anomaly in the heat capacity, as shown in Figure 7 (for x=0.15). The superconducting contribution, defined as the difference between the superconducting and normal state $C_p$'s, i.e. $\Delta C_p(H) = C_p(H) - C_p(30\text{ kOe})$, is clearly resolved by subtracting the heat capacity data measured in magnetic fields above $H_{c2}$ (see inset Figure 7). The $C_p$-peak shifts to lower temperature with increasing field, as expected for a superconducting transition. The jump of the heat capacity, $\Delta C_p/T_c$ = 8 mJ mol$^{-1}$ K$^{-2}$, is comparable to values reported for doped iron arsenides.[23]

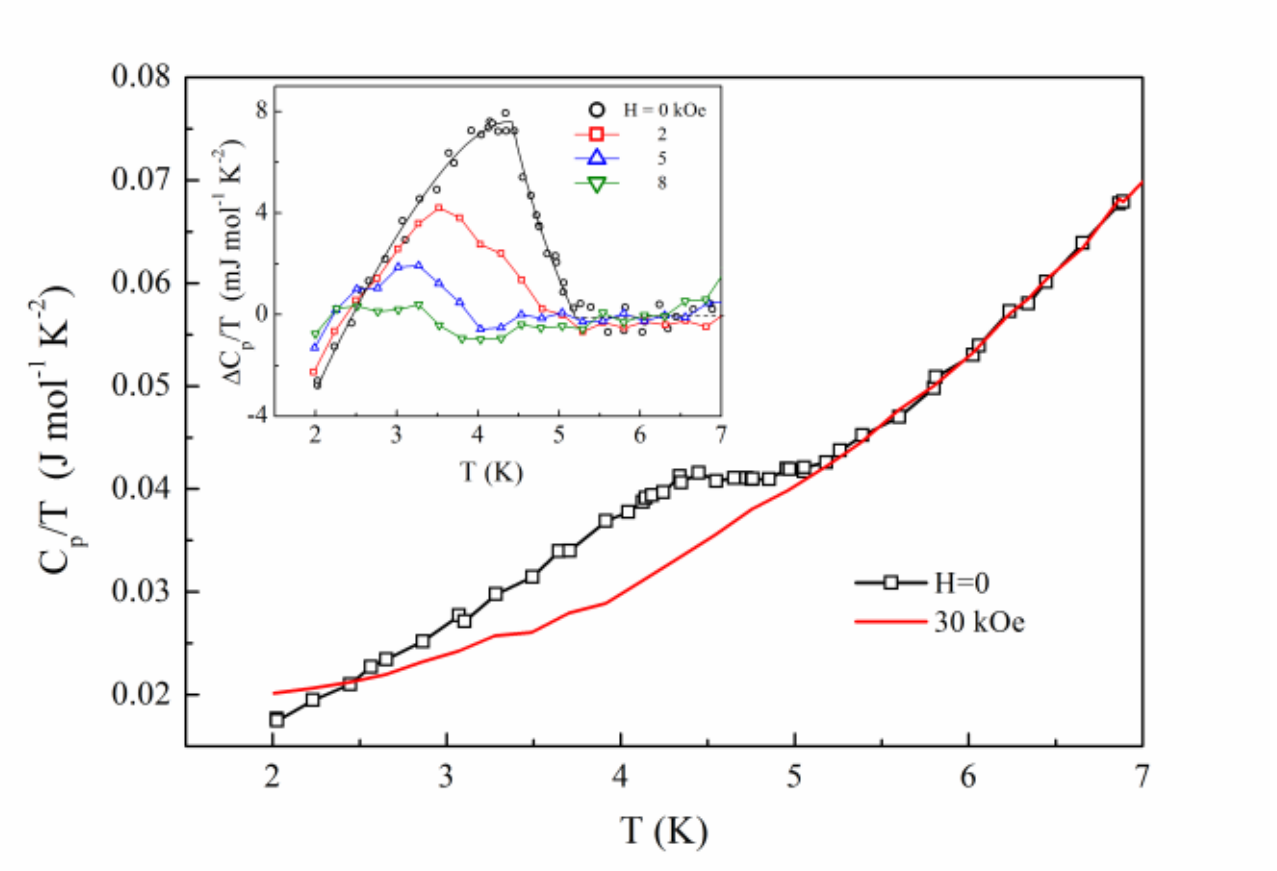

***Figure 7.*** Low-temperature heat capacity of $Ba_{0.85}Na_{0.15}Ti_2Sb_2O$. The inset shows the difference between the heat capacities of the superconducting and normal states.

The heat capacity and magnetization data with a large diamagnetic signal provide convincing evidence that a bulk superconducting state is achieved in $BaTi_2Sb_2O$ with Na substitution. Thus, we have found a new layered Ti-based pnictide oxide wherein a superconducting state arises from a magnetically ordered state when properly doped, akin to the HTS cuprates and iron pnictides.

Although the transition temperatures in Na-doped $BaTi_2Sb_2O$ are relatively low, there are significant and striking similarities with the cuprate and iron pnictide superconductors. The $Ti_2O$ square nets in $BaTi_2Sb_2O$ are inverse to the $CuO_2$ planes in the cuprates. Moreover, Ti is formally trivalent with $d^1$ configuration, analogous to the "hole" $d^9$ configuration of Cu in the cuprates. In addition, the Sb atoms, bonded to Ti atoms above and below the $Ti_2O$ planes, significantly contribute to the magnetic superexchange between the Ti spins,[12] which resembles the electronic situation in the superconducting iron pnictides. The existence of a nesting property of the Fermi surface of undoped pnictide oxide compounds[12] and the associated CDW/SDW instability accompanied by a structural distortion[14] is another feature that has been observed in the iron pnictides. More importantly superconductivity emerges from the magnetically ordered state upon doping, and the critical temperature of the CDW/SDW instability ($T_s$) is suppressed to lower temperatures. However, in contrast to the iron pnictides, the CDW/SDW state survives at relatively higher doping levels (Figure 5 inset). Unfortunately, the solubility of Na in $BaTi_2Sb_2O$ was limited at ~33% and the phase diagram could not be extended beyond this limit. The softening of the resistivity anomaly at $T_S$ at higher Na content (Figure 5) may also indicate an abrupt disappearance of the CDW/SDW phase above a critical doping, similar to the phase diagram of $LaO_{1-x}F_xFeAs$.[24]

In view of the unsuccessful attempts to induce superconductivity in $BaTi_2As_2O$[17] and the relatively low Na content (5 %) to induce the superconducting state in $BaTi_2Sb_2O$, a question arises about the major differences between doping in $BaTi_2As_2O$ and in $BaTi_2Sb_2O$. One significant difference of both parent compounds is the much lower energy scale of the CDW/SDW phase in $BaTi_2Sb_2O$ resulting in the lowest critical temperature ($T_s$ = 54 K) in all Ti-based pnictide oxides. Since the competition of magnetic order and superconductivity is a crucial element in unconventional superconductivity, it is conceivable that this lower energy scale is the key to what makes superconductivity to emerge upon doping with Na in $BaTi_2Sb_2O$. Further work will focus on exploring the detailed phase diagram of the new superconductors, and the nature of the CDW/SDW phase and of the emerging superconductivity in $BaTi_2Sb_2O$. Exploratory synthesis of other layered Ti-based pnictide oxide superconductors and related compounds may yet lead to new families of unconventional superconductors.

## ACKNOWLEDGMENT.

This work is supported in part by the State of Texas through the Texas Center for Superconductivity, US Air Force Office of Scientific Research, National Science Foundation (CHE-0616805), and the R.A. Welch Foundation (E-1297). P.C.W.C. acknowledges the T.L.L. Temple Foundation, the J.J. and R. Moores Endowment, and LBNL through the USDOE.

### Supporting Information Available:

Experimental procedures, results of the elemental analyses, summary of crystallographic data, and a list of lattice parameters and their trends are included in the Supporting Information. This material is available free of charge via the Internet at http://pubs.acs.org.

Corresponding Author:
Prof. Arnold M. Guloy
136 Fleming Bldg., Department of Chemistry
University of Houston, Houston, TX 77204-5003
email: aguloy@uh.edu

# $Ba_{1-x}Na_xTi_2Sb_2O$ (0.0 ≤ x≤ 0.33): A Layered Titanium-based Pnictide Oxide Superconductor.

Phuong Doan[†,Π], Melissa Gooch[‡,Π], Zhongjia Tang[†,Π], Bernd Lorenz[‡,Π], Angela Möller [†,Π], Joshua Tapp[†,Π], Paul C.W. Chu[‡,Π,§] and Arnold M. Guloy[*,†,Π].

[†]Department of Chemistry, [‡]Department of Physics, and [Π]Texas Center for Superconductivity, University of Houston, Houston, TX 77204; [§]Lawrence Berkeley National Laboratory, 1 Cyclotron Road, Berkeley, CA 94720.

**SUPPORTING INFORMATION:**

# I. Experimental Procedures

All manipulations and weighings were performed in an Ar-filled glovebox (total $O_2$ + $H_2O$ < 0.1 ppm) to ensure that the total content of each reaction was known well and that minimum contamination occurred during the handling of the reagents. As is customary, the 1/4-in o.d. Nb containers utilized showed no visible attack and retained their ductility through the reactions. Mixtures of the reagents therein were based on a 300-500 mg scale.

**Synthesis of $BaTi_2Sb_2O$**

$BaTi_2Sb_2O$ was synthesized by reacting stoichiometric amounts of BaO (99.99% Aldrich) and Ti (99.99% Aldrich), Sb pieces (99.999%, Alfa Aesar ) in welded Nb containers within evacuated quartz jackets. The reactants were mixed, ground, then pressed into pellets. These were sealed in Nb (1/4" diameter) tubes under argon, and subsequently jacketed, evacuated and sealed in quartz tubes. The samples were heated slowly (2ºC/min) to 900ºC and annealed for 3 days, then slowly cooled (1 ºC/min) to 200ºC. Additional regrinding and sintering at 900ºC for another 3 days was performed to ensure phase homogeneity. The air- and moisture sensitive polycrystalline product had a dark gray color.

**Synthesis of Na-doped $Ba_{1-x}Na_xTi_2Sb_2O$**

$Ba_{1-x}Na_xTi_2Sb_2O$ were synthesized by reacting stoichiometric amounts of BaO (99.99%, Sigma Aldrich); $BaO_2$ (95% , Sigma Aldrich); $Na_2O$ (80% $Na_2O$ and 20% $Na_2O_2$, Sigma Aldrich); Ti (99.99% , Sigma Aldrich), Sb (pieces, 99.999%, Alfa Aesar) in welded Nb containers within evacuated quartz jackets. The major impurity in $Na_2O$ (20% $Na_2O_2$) had to be taken into account during the weighing. The reactants were mixed, ground and pressed into pellets, then pressed into pellets. These were sealed in Nb (1/4" diameter) tubes under argon, and subsequently jacketed, evacuated and sealed in quartz tubes. The samples was heated slowly (2ºC/min) to 900ºC and annealed for 3 days, then slow cooled (1 ºC/min) to 200ºC. Additional regrinding and sintering at 900ºC for another 3 days was performed to ensure phase homogeneity. The air- and moisture sentsitive polycrystalline product had a dark gray color.

**Details of the Physical Measurements:**

Magnetic measurements on small compressed pellets were performed using a Magnetic Property Measurement System (MPMS, Quantum Design). Resistivity measurements on the sample pellets were carried out using on a home-made resistivity puck with attaching wires using indium contacts. The measurements were conducted using the low-frequency resistance bridge (LR700) in combination with the Physical Property Measurement System (PPMS, Quantum Design) which provided the temperature and magnetic field control. Heat capacity was measured using a relaxation method of the PPMS.

## II. Results of elemental analyses by ICP-MS

The elemental analysis was carried out using an ICP-MS system from Varian. Introduction of the sample into the ICPMS were made via laser ablation (LSX-213 Laser Ablation System) on selected crystallites and bulk pieces from XRD phase pure samples. Results of the elemental analyses for Ba and Na, scaled with respect to Ti content, are reported below.

| | Na Nominal Composition (x) | | | | | |
|---|---|---|---|---|---|---|
| $Ba_{1-x}Na_xTi_2Sb_2O$ | 0.00 | 0.05 | 0.10 | 0.25 | 0.30 | 0.33 |
| Na analyses, (x) | 0 | 0.056(1) | 0.091(1) | 0.24(3) | 0.295(9) | 0.33(2) |
| Ba analyses (1-x) | 1.03(3) | 0.93(1) | 0.88(1) | 0.72(1) | 0.68(3) | 0.62(2) |
| Ti analyses (set =2) | 2 | 2 | 2 | 2 | 2 | 2 |

## III. XRD Powder data (with indices) and summary of the Rietveld Refinement for $BaTi_2Sb_2O$

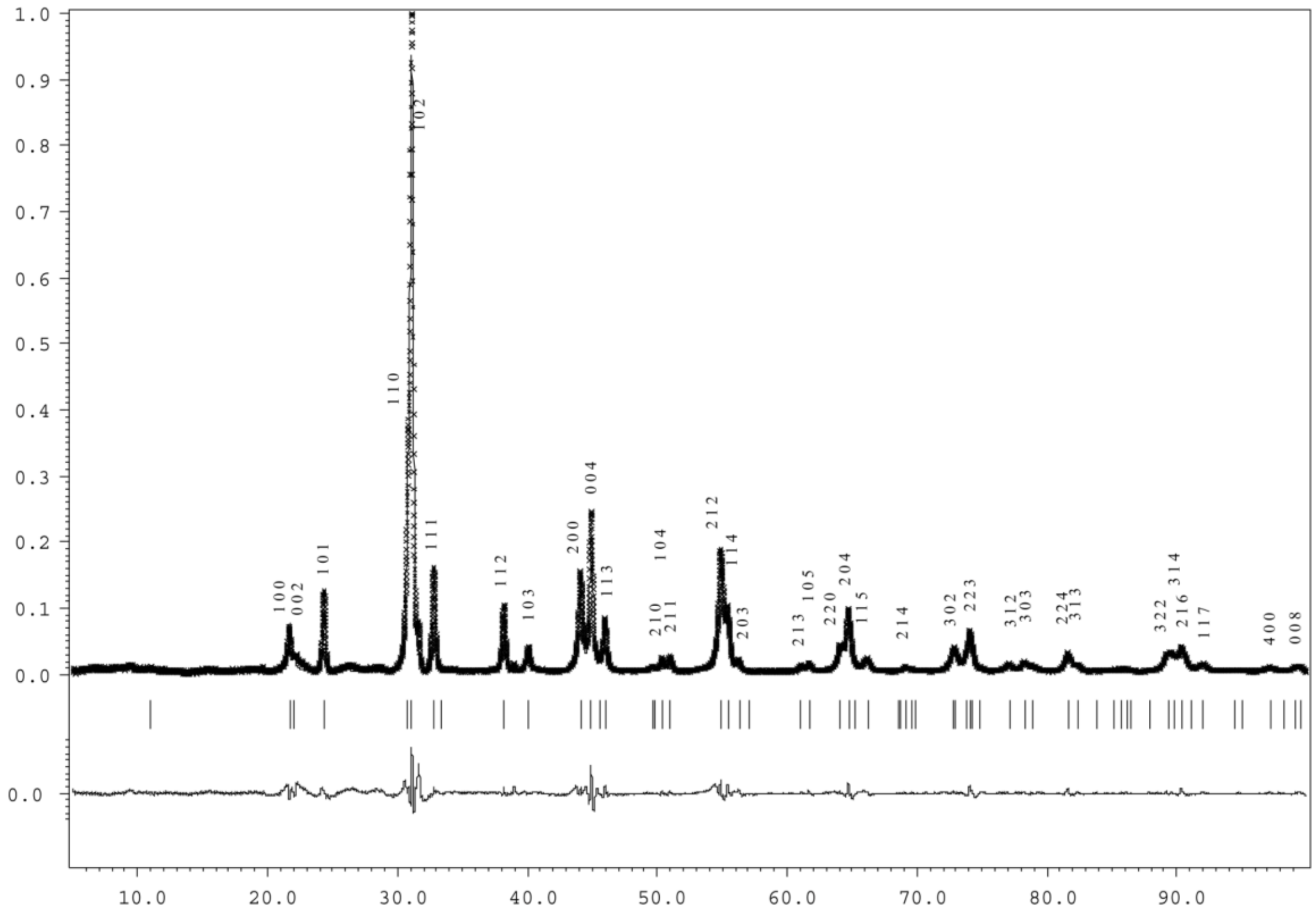

# Results of Rietveld refinement of $BaTi_2Sb_2O$ XRD data

- $CeCr_2Si_2C$-type; space group P 4/mmm
- $a$ = 4.1196(1); $c$ = 8.0951(2)Å, Vol = 137.383(6), $D_{calc}$= 6.003g/cm$^3$
- Rp = 6.30%; Rwp = 5.05%

| Atomic parameters | | | | | | | | |
|---|---|---|---|---|---|---|---|---|
| **Atom** | **Ox.** | **Wyck.** | **Site** | **S.O.F.** | **x/a** | **y/b** | **z/c** | **U [Å²]** |
| Ba | | 1d | 4/mmm | 1.0 | 0 | 0 | 0 | 0.0089(7) |
| Ti | | 2f | mmm. | 1.0 | 0 | 0.5 | 0.5 | 0.0102(1) |
| Sb | | 2g | 4mm | 1.0 | 0.5 | 0.5 | 0.2514(1) | 0.0049(6) |
| O | | 1c | 4/mmm | 1.0 | 0 | 0 | 0.5 | 0.124 (8) |

## IV. Relevant Bond lengths [Å] for $BaTi_2Sb_2O$

- Ti-O 2.054(1)Å
- Ti-Sb 2.872(1)Å
- Ti-Ti 2.905(1)Å

- Ba---Sb 3.543(2)Å
- Sb---Sb 4.1076(1)Å ( *a*),
- 4.0592(2)Å ( *c*)
- 4.0139(2)Å ( *c*).

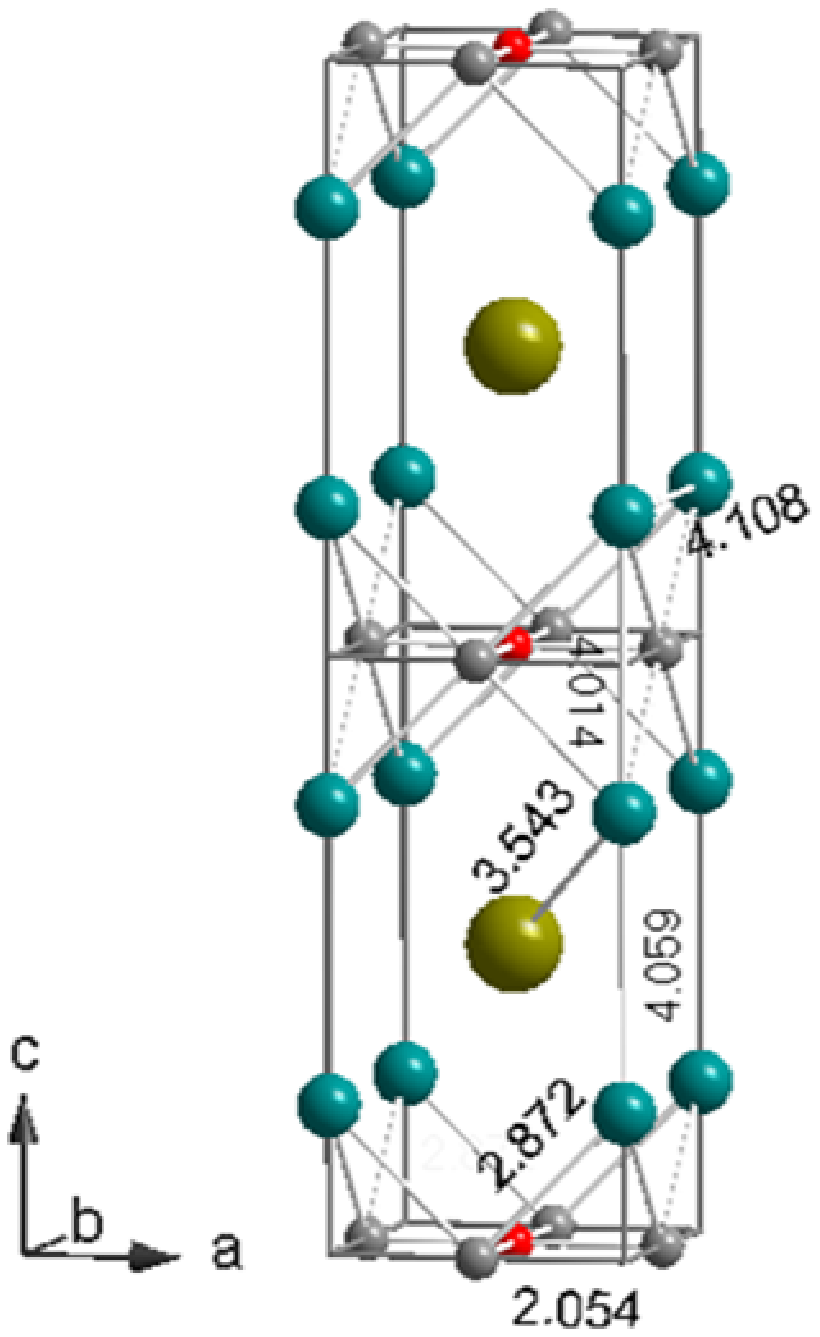

## V. Table of Unit Cell (Lattice) parameters of $Ba_{1-x}Na_xTi_2Sb_2O$ and their trends

| | *a* | *c* | Volume |
|---|---|---|---|
| $BaTi_2Sb_2O$ | 4.1196(1) | 8.0951(2) | 137.35 |
| $Ba_{0.95}Na_{0.05}Ti_2Sb_2O$ | 4.1111(2) | 8.097(2) | 136.85 |
| $Ba_{0.90}Na_{0.10}Ti_2Sb_2O$ | 4.1091(4) | 8.0937(8) | 136.66 |
| $Ba_{0.85}Na_{0.15}Ti_2Sb_2O$ | 4.1008(5) | 8.092(1) | 136.07 |
| $Ba_{0.80}Na_{0.20}Ti_2Sb_2O$ | 4.0969(5) | 8.096(1) | 135.89 |
| $Ba_{0.75}Na_{0.25}Ti_2Sb_2O$ | 4.0944(4) | 8.105(1) | 135.88 |
| $Ba_{0.70}Na_{0.30}Ti_2Sb_2O$ | 4.0909(5) | 8.106(1) | 135.66 |
| $Ba_{0.67}Na_{0.33}Ti_2Sb_2O$ | 4.0932(5) | 8.113(2) | 135.93 |
| $Ba_{0.65}Na_{0.35}Ti_2Sb_2O$ | 4.090(1) | 8.111(3) | 135.67 |

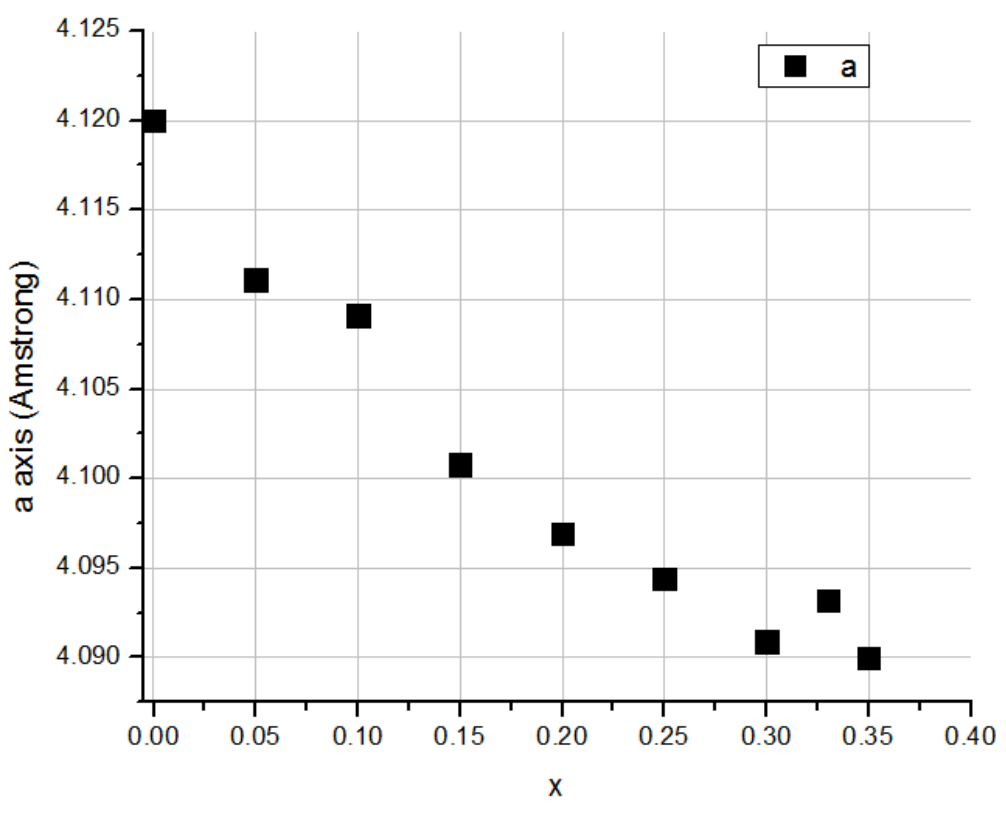

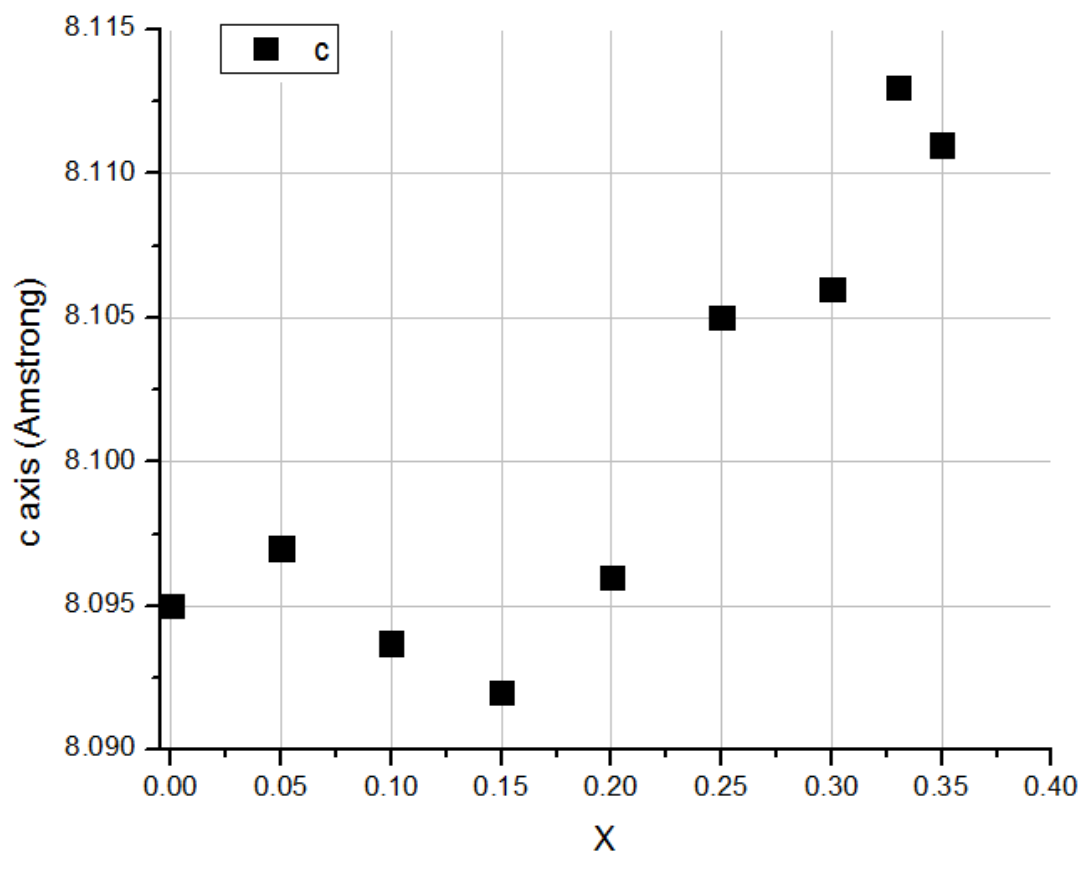

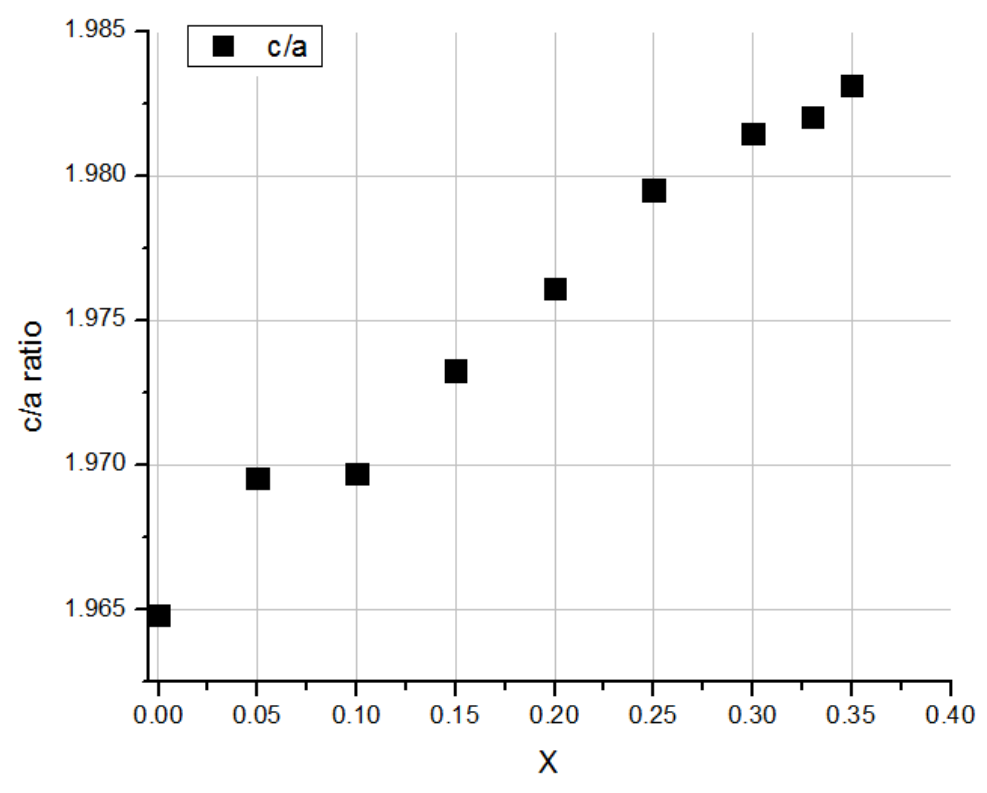

**VI. Phase Diagram of the CDW/SDW – Superconductivity of C $Ba_{1-x}Na_xTi_2Sb_2O$ (as enlarged from the inset in Figure 5)**

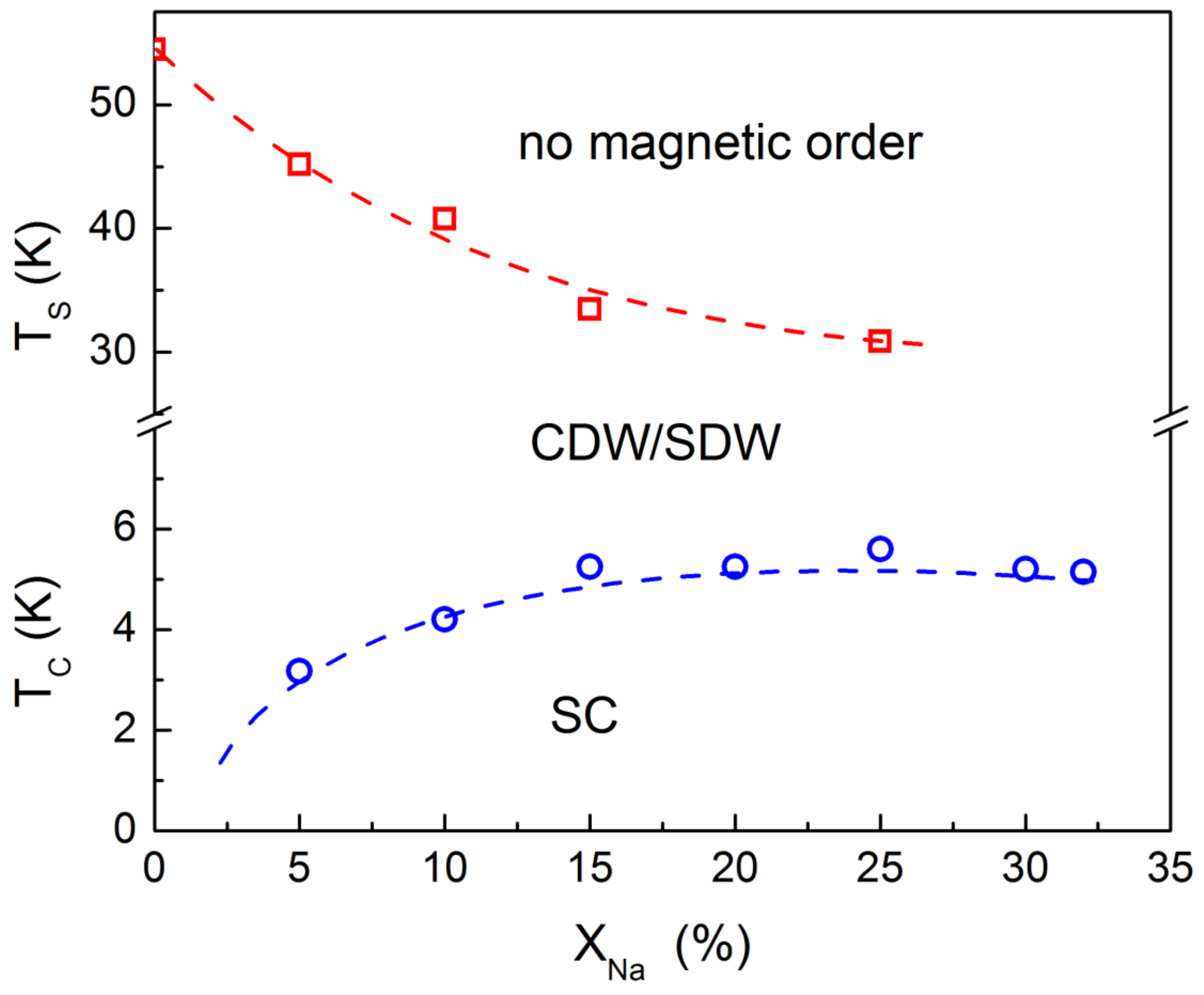